# Mid-IR and VUV Spectroscopic Characterisation of Thermally Processed and Electron Irradiated CO$_2$ Astrophysical Ice Analogues

D.V. Mifsud[1,2,*], Z. Kaňuchová[3,4,*], S. Ioppolo[5], P. Herczku[2], A. Traspas Muiña[5], T.A. Field[6], P.A. Hailey[1], Z. Juhász[2], S.T.S. Kovács[2], N.J. Mason[1], R.W. McCullough[6], S. Pavithraa[7,†], K.K. Rahul[7,‡], B. Paripás[8], B. Sulik[2], S.-L. Chou[9], J.-I. Lo[9,l], A. Das[10], B.-M. Cheng[9,l], B.N. Rajasekhar[11], A. Bhardwaj[7], and B. Sivaraman[7,*]

[1] *Centre for Astrophysics and Planetary Science, School of Physical Sciences, University of Kent, Canterbury CT2 7NH, United Kingdom*

[2] *Atomic and Molecular Physics Laboratory, Institute for Nuclear Research (Atomki), Debrecen H-4026, Hungary*

[3] *Astronomical Institute, Slovak Academy of Sciences, Tatranska Lomnicá SK-059 60, Slovakia*

[4] *INAF Osservatorio Astronomico di Roma, Monte Porzio Catone RM-00078, Italy*

[5] *School of Electronic Engineering and Computer Science, Queen Mary University of London, London E1 4NS, United Kingdom*

[6] *Department of Physics and Astronomy, School of Mathematics and Physics, Queen's University Belfast, Belfast BT1 7NN, United Kingdom*

[7] *Atomic, Molecular, and Optical Physics Division, Physical Research Laboratory, Ahmedabad 380 009, India*

[8] *Department of Physics, Faculty of Mechanical Engineering and Informatics, University of Miskolc, Miskolc H-3515, Hungary*

[9] *National Synchrotron Radiation Research Centre (NSRRC), Hsinchu 300, Taiwan*

[10] *Department of Astrochemistry and Astrobiology, Indian Centre for Space Physics, Kolkata 700 084, India*

[11] *Atomic and Molecular Physics Division, Bhabha Atomic Research Centre, Mumbai 400 085, India*

[†] Current Affiliation: *Institute of Molecular Science, National Chiao Tung University, Hsinchu 30010, Taiwan*

[‡] Current Affiliation: *Atomic and Molecular Physics Laboratory, Institute for Nuclear Research (Atomki), Debrecen H-4026, Hungary*

[l] Current Affiliation: *Department of Medical Research, Hualien Tzu Chi Hospital, Buddhist Tzu Chi Medical Foundation, Hualien City 970, Taiwan*

\* Corresponding Authors:  
    D.V. Mifsud    duncanvmifsud@gmail.com  
    Z. Kaňuchová    pipovci@gmail.com  
    S. Ioppolo    s.ioppolo@qmul.ac.uk  
    B. Sivaraman    bhala@prl.res.in

**ORCID Identification Numbers**

- D.V. Mifsud        0000-0002-0379-354X
- Z. Kaňuchová       0000-0001-8845-6202
- S. Ioppolo         0000-0002-2271-1781
- P. Herczku         0000-0002-1046-1375
- A. Traspas Muiña   0000-0002-4304-2628
- T.A. Field         0000-0002-3394-1136
- P.A. Hailey        0000-0002-8121-9674
- Z. Juhász          0000-0003-3612-0437
- S.T.S. Kovács      0000-0001-5332-3901
- N.J. Mason         0000-0002-4468-8324
- R.W. McCullough    0000-0002-4361-8201
- S. Pavithraa       0000-0001-8809-7633
- K.K. Rahul         0000-0002-5914-7061
- B. Paripás         0000-0003-1453-1606
- B. Sulik           0000-0001-8088-5766
- S.-L. Chou         0000-0002-6528-3721
- J.-I. Lo           0000-0002-0153-1423
- A. Das             0000-0003-4615-602X
- B.-M. Cheng        0000-0002-8540-6274
- B.N. Rajasekhar    0000-0002-0038-8768
- A. Bhardwaj        0000-0003-1693-453X
- B. Sivaraman       0000-0002-2833-0357

# Mid-IR and VUV Spectroscopic Characterisation of Thermally Processed and Electron Irradiated $CO_2$ Astrophysical Ice Analogues

The astrochemistry of $CO_2$ ice analogues has been a topic of intensive investigation due to the prevalence of $CO_2$ throughout the interstellar medium and the Solar System, as well as the possibility of it acting as a carbon feedstock for the synthesis of larger, more complex organic molecules. In order to accurately discern the physico-chemical processes in which $CO_2$ plays a role, it is necessary to have laboratory-generated spectra to compare against observational data acquired by ground- and space-based telescopes. A key factor which is known to influence the appearance of such spectra is temperature, especially when the spectra are acquired in the infrared and ultraviolet. In this present study, we describe the results of a systematic investigation looking into: (i) the influence of thermal annealing on the mid-IR and VUV absorption spectra of pure, unirradiated $CO_2$ astrophysical ice analogues prepared at various temperatures, and (ii) the influence of temperature on the chemical products of electron irradiation of similar ices. Our results indicate that both mid-IR and VUV spectra of pure $CO_2$ ices are sensitive to the structural and chemical changes induced by thermal annealing. Furthermore, using mid-IR spectroscopy, we have successfully identified the production of radiolytic daughter molecules as a result of 1 keV electron irradiation and the influence of temperature over this chemistry. Such results are directly applicable to studies on the chemistry of interstellar ices, comets, and icy lunar objects and may also be useful as reference data for forthcoming observational missions.

## 1    Introduction

Amongst the various molecules known to populate interstellar icy grain mantles, $CO_2$ is one of the more prevalent species having an estimated abundance of 20-30% relative to that of $H_2O$ (Öberg 2016). $CO_2$ ice is also widely distributed within the Solar System, perhaps most famously within the polar ice caps on Mars (Phillips et al. 2011, Manning et al. 2019), as well as in cometary material (Filacchione et al. 2016, Läuter et al. 2018).

$CO_2$ ice has also been detected as a surface component of several outer Solar System moons, including the icy satellites of Jupiter, Saturn, and Uranus (e.g., McCord et al. 1997, McCord et al. 1998, Hibbitts et al. 2000, Hibbitts et al. 2003, Buratti et al. 2005, Grundy et al. 2006, Cartwright et al. 2015, Sori et al. 2017, Combe et al. 2019). Given this cosmic ubiquity, it is no surprise that the spectroscopy and chemistry of $CO_2$ ice have been the subject of intensive investigation.

Within interstellar space, the vast majority of $CO_2$ is thought to be formed directly in the solid phase, as modelled gas-phase processes are not able to reproduce the abundances observed (Boonman et al. 2003). This formation results from the neutral-neutral reaction between condensed CO and OH which proceeds via a HO–CO intermediate complex (Goumans et al. 2008, Ioppolo et al. 2011). Alternatively, the energetic processing of ices rich in carbon- and oxygen-containing molecules by galactic cosmic rays or UV photons, or of $H_2O$ ice deposited on carbonaceous dust grains, may contribute to the interstellar abundance of $CO_2$ ice, as has been demonstrated by several laboratory studies (e.g., Moore et al. 1991, Mennella et al. 2004, Loeffler et al. 2005, Gomis and Strazzulla 2005, Mennella et al. 2006, Ioppolo et al. 2009, Jheeta et al. 2013, Mason et al. 2014, Arumainayagam et al. 2019).

Interstellar $CO_2$ ices may exist as amorphous structures within polar phases of the icy grain mantles which are dominated by molecules such as $H_2O$, or in apolar phases alongside other molecules such as CO and $O_2$. Under the influence of thermal annealing, pure $CO_2$ may form as a result of segregation from the polar phase or the distillation of volatile molecules in the apolar phase (Escribano et al. 2013, He et al. 2017). Studies have shown that this pure $CO_2$ ice exhibits at least some form of structural order, such as polycrystallinity (Escribano et al. 2013, Allodi et al. 2014, McGuire et al. 2016, Tsuge et al. 2020, Kouchi et al. 2021). Additionally, it appears that this crystallinity persists at the low temperatures encountered within dense interstellar molecular clouds and is resistant to UV photolysis or particle radiolysis (Hudson and Moore 1995, Tsuge et al. 2020).

Spectroscopic studies of interstellar $CO_2$ made using space-borne observatories such as, for example, the *Infrared Space Observatory* and the *Spitzer Space Telescope*, have been used extensively to infer information on icy grain mantle compositions, structures, and temperatures. The infrared spectroscopic features of $CO_2$ are especially sensitive to these parameters (Sandford and Allamandola 1990, Bernstein et al. 2005, Isokoski et al. 2013). In the solid phase, $CO_2$ presents four absorption bands in the mid-infrared region (Gerakines et al. 1999): an asymmetric stretching mode ($v_3$) at 2342 $cm^{-1}$ (4.27 μm), a bending mode ($v_2$) at 660 $cm^{-1}$ (15.20 μm), and two combination modes ($v_1 + v_3$ and $2v_2 + v_3$) located at 3708 $cm^{-1}$ (2.70 μm) and 3600 $cm^{-1}$ (2.78 μm). The symmetric stretching mode ($v_1$) is infrared inactive as it does not produce a variation in the molecular dipole moment.

The $v_2$ bending mode has traditionally proven particularly useful in gleaning information on the chemical (i.e., molecular) environment of $CO_2$ (e.g., d'Hendecourt and Jourdain de Muizon 1989, Pontoppidan et al. 2008, Cook et al. 2011). It has been claimed that, in amorphous ices, this band presents as a broad single-peaked structure while a double-peaked structure is observed in pure crystalline ices (Gerakines and Hudson 2015). However, more recent work has disputed this, and has suggested that the appearance of this band as a single-peaked structure is more likely due to the presence of contaminant species, such as $H_2O$, in the ice matrix rather than a difference in phase (Baratta and Palumbo 2017). Alternative spectral indicators of phase, such as Fermi resonances in the amorphous phase (Amat and Pimbert 1965) and Davydov splitting in the crystalline phase (Davydov and Sardaryan 1962), have been suggested. He and Vidali (2018) have also suggested that the $v_1 + v_3$ combination mode may be used to quantify the degree of crystallinity in thin $CO_2$ films. The structure of the $\nu_3$ stretching mode can also provide information on ice morphology. For instance, Escribano et al. (2013) noted that a 'shoulder' band was apparent at 2328 cm$^{-1}$ (4.30 μm) in the pure amorphous phase which was not visible upon crystallisation via thermal annealing. This band has also not been detected in astronomical observations, thus implying at least some degree of structural order in pure interstellar $CO_2$ ice analogues (Allodi et al. 2014, McGuire et al. 2016, Tsuge et al. 2020).

Although perhaps less popular than infrared spectroscopy, electronic spectroscopy has also been used in astrochemical studies of $CO_2$ ice analogues (e.g., Mason et al. 2006, Pavithraa et al. 2019, James et al. 2019, James et al. 2020). Between 120 nm (10.3 eV) and 180 nm (6.9 eV), $CO_2$ ice presents two broad absorption bands centred about 125 nm (9.9 eV) and 141 nm (8.8 eV) corresponding to the $^1\Pi_g \leftarrow {}^1\Sigma_g^+$ and the $^1\Delta_u \leftarrow {}^1\Sigma_g^+$ transitions, respectively. These transitions arise due to photon-induced promotion of the $1\pi_g$ electron to the $3s\sigma_g$ orbital and to the first unoccupied $2\pi_u^*$ orbital, respectively.

Previous work has established that the absorption spectra of $CO_2$ astrophysical ice analogues vary somewhat in their appearance depending on the temperature of the ices. For instance, Isokoski et al. (2013) performed an analysis of highly resolved mid-infrared spectra of solid $CO_2$ at various temperatures in the range 15-75 K at 15 K intervals. This work allowed them to determine that the $\nu_3$ stretching mode red-shifts to marginally lower wavenumbers and becomes narrower and sharper in appearance on warming to higher temperatures due to molecular ordering into a more crystalline structure and the associated reduction in molecular environments. Additionally, they noted that the $\nu_2$ bending mode presented as a double-peaked structure with the higher wavenumber peak exhibiting a 'wing' structure which declined under the influence of thermal annealing, in line with other studies (e.g., Sandford and Allamandola 1990, Ehrenfreund et al. 1997, van Broekhuizen et al. 2006).

Electronic absorption spectra of solid $CO_2$, on the other hand, are not as common in the literature, with only a handful of studies available (e.g., Monahan and Walker 1974, Deschamps et al. 2003, Mason et al. 2006). Indeed, to the best of the authors' knowledge, there is no report currently available where the primary focus is the thermal evolution of the electronic absorption spectra of $CO_2$ ice. The work of James et al. (2020) did make mention of such a processing, but was more focused on the thermal processing of mixed $CO_2$:$NH_3$ ices and the resultant chemistry as monitored in the vacuum-ultraviolet. As such, there currently exists a gap in the literature in this regard.

Temperature may also be an influential factor in the chemistry resulting from the irradiative processing of $CO_2$ ices. Interstellar and Solar System ices are routinely exposed to irradiation by UV photons or charged particles (the latter from galactic cosmic rays or the solar wind), and laboratory studies have shown that such processing results in the formation of a number of daughter molecules, the most prominent of which are CO, $CO_3$, and $O_3$ (e.g., Palumbo et al. 1997, Satorre et al. 2000, Hudson and Moore 2001, Sivaraman et al. 2013, Martín-Doménech et al. 2015). These products have also been observed after the irradiation of mixed ices containing $CO_2$, along with other molecules. For example, the irradiation of mixed $CO_2$:$H_2O$ ices additionally produces $H_2O_2$ and $H_2CO_3$ (e.g., Pirronello et al. 1982, Moore and Khanna 1991, Brucato et al. 1997, Wu et al. 2003, Strazzulla et al. 2005, Pilling et al. 2010, Radhakrishnan et al. 2018, Pavithraa et al. 2019). Perhaps most interesting is the fact that the irradiation of multi-component ices containing $CO_2$ has been shown to result in the formation of complex organic and prebiotic molecules such as chiral amino acids (Muñoz-Caro et al 2002, Hudson et al. 2008, Esmaili et al. 2018).

The aim of this present study, therefore, is two-fold: (i) to present a systematic and coherent characterisation of the influence of temperature on the appearance of the mid-infrared (mid-IR) and vacuum-ultraviolet (VUV) absorption spectra of pure $CO_2$ astrophysical ice analogues, and (ii) to quantify the effect of temperature on the production of radiolytic daughter molecules as a result of the irradiation of such ices using high-energy electrons. To achieve these aims, we have measured firstly the mid-IR and VUV spectra of pure, unirradiated $CO_2$ deposited at 10 (VUV only), 25, 30, 40, and 50 K as they are thermally annealed to 70 K. Secondly, we have irradiated pure $CO_2$ ices deposited at 20, 30, 40, and 50 K using 1 keV electrons and monitored the resultant chemistry using mid-IR spectroscopy.

## 2    Materials and Methods

The work presented in this study is the result of a multi-national collaborative effort making use of experimental set-ups at two large-scale facilities: (i) the Ice Chamber for Astrophysics-Astrochemistry (ICA) at the Institute for Nuclear Research (Atomki) in Debrecen, Hungary, and (ii) the photo-absorption and photo-luminescence end-station at the high-flux BL 03 beam line on the Taiwan Light Source (TLS) at the National Synchrotron Radiation Research Centre (NSRRC) in Hsinchu, Taiwan. In this section, a description of the apparatus and experimental protocols used at each facility is provided.

### 2.1    *Experiments Performed at Atomki*

The experimental work performed at Atomki made use of the ICA set-up, which has been described in great detail elsewhere (Herczku et al. 2021, Mifsud et al. 2021a). Briefly, the ICA is a high-vacuum chamber containing a series of ZnSe substrates vertically mounted onto a copper sample holder, which is cooled to 20 K by a closed-cycle helium cryostat. The temperature of the substrates is measured using two silicon diodes and can be regulated over the range 20-300 K using a temperature controller system. The pressure in the chamber is typically maintained at a few $10^{-9}$ mbar by the combined use of a dry rough vacuum pump and a turbomolecular pump. Astrophysical ice analogues are accreted onto the cold ZnSe windows via background deposition of gases and vapours which are introduced into a mixing chamber prior to their dosing into the main chamber through an all-metal needle valve.

Ices are monitored *in situ* during deposition and processing using mid-IR transmission absorption spectroscopy. A Thermo-Nicolet Nexus 670 Fourier-transform spectrophotometer with a spectral range of 4000-650 cm$^{-1}$ and a resolution of 1 cm$^{-1}$ with a mercury-cadmium-telluride external detector was used for this study, and allowed for a quantitative determination of the thickness $d$ (μm) of the deposited ices. This was done by first calculating the molecular column density $C$ (molecules cm$^{-2}$) of the deposited $CO_2$ ice by integrating the Beer-Lambert Equation (Eq. 1) over the $v_3$ absorption band (Herczku et al. 2021). Once calculated, the column density was used to determine the ice thickness as per Eq. 2:

$$C = \frac{1}{A_v} \int \tau(v)\, dv$$

(Eq. 1)

$$d = \frac{CZ}{\rho N_A} \times 10^4$$

(Eq. 2)

where $A_v$ is the integrated band strength which we have taken to be 7.6×10$^{-17}$ cm molecule$^{-1}$ (Gerakines et al. 1995), $\tau(v)$ is the optical depth (cm$^{-1}$), $Z$ is the molecular mass of the $CO_2$ molecule (44 g mol$^{-1}$), $\rho$ is the density of the ice which we have taken to be 0.98 g cm$^{-3}$ (Luna et al. 2012), and $N_A$ is the Avogadro constant (6.02×10$^{23}$ molecules mol$^{-1}$).

Two aspects of work were carried out at Atomki using the ICA set-up, both of which relying on mid-IR transmission absorption spectroscopy as the primary analytical method. Firstly, thermal annealing studies of pure, unirradiated $CO_2$ ices initially deposited at 25, 30, 40, and 50 K were performed so as to quantify the influence of temperature on the resultant mid-IR spectra. Secondly, pure $CO_2$ ices were irradiated at 20, 30, 40, and 50 K using 1 keV electrons so as to observe the effect of temperature on the formation of radiolytic product species, particularly CO, $CO_3$, and $O_3$.

For the thermal annealing experiments, reference spectra of the bare ZnSe substrates were collected in the temperature range 25-70 K at 5 K intervals. The substrates were then cooled to the desired initial temperature (i.e., 25, 30, 40, or 50 K) for deposition of the pure $CO_2$ ice. Gaseous $CO_2$ (99.995% from Linde MINICAN) was pre-prepared in a mixing chamber, where its pressure was determined using standard manometric practices. The ice was then prepared via background deposition at a chamber pressure of a few $10^{-6}$ mbar. In these experiments, pure $CO_2$ ices with thicknesses of 0.04-0.07 μm were prepared. Once the ice was deposited, a mid-IR absorption spectrum was collected at the deposition temperature, and the ice was subsequently warmed to 70 K with further spectra being acquired every 5 K using the appropriate reference spectrum.

The second aspect of the work at Atomki involved electron irradiation of pure $CO_2$ ices at 20, 30, 40, and 50 K. The pure ices were prepared as described above and deposited to thicknesses of 0.45-0.55 μm. Once the ice was deposited at the desired temperature, a mid-IR absorption spectrum was acquired and the deposited samples were irradiated with 1 keV electrons at an incidence angle of 36° to the normal with additional spectra collected as required. The electron beam current and profile had been measured before commencing the experiment using the method described by Mifsud et al. (2021a), and the total fluence delivered $\varphi$ (electrons cm$^{-2}$) was found using Eq. 3:

$$\varphi = \int_0^t \frac{I}{Ae}\,dt$$

(Eq. 3)

where *I* is the beam current, *e* is the fundamental electric charge ($1.602\times10^{19}$ C), *A* is the area of the substrate exposed to the electron beam (which, in this set-up, was 1.13 cm$^2$), and *t* is the time of irradiation. The final fluence delivered in this set of experiments varied between $5.3\text{-}6.9\times10^{16}$ electrons cm$^{-2}$. The penetration depth of the impinging electrons was also estimated using the CASINO software (Drouin et al. 2007) and was found to 0.07 μm (Fig. 1). Given that this penetration depth is smaller than the estimated ice thickness, all of the electrons were effectively implanted into the CO$_2$ ice.

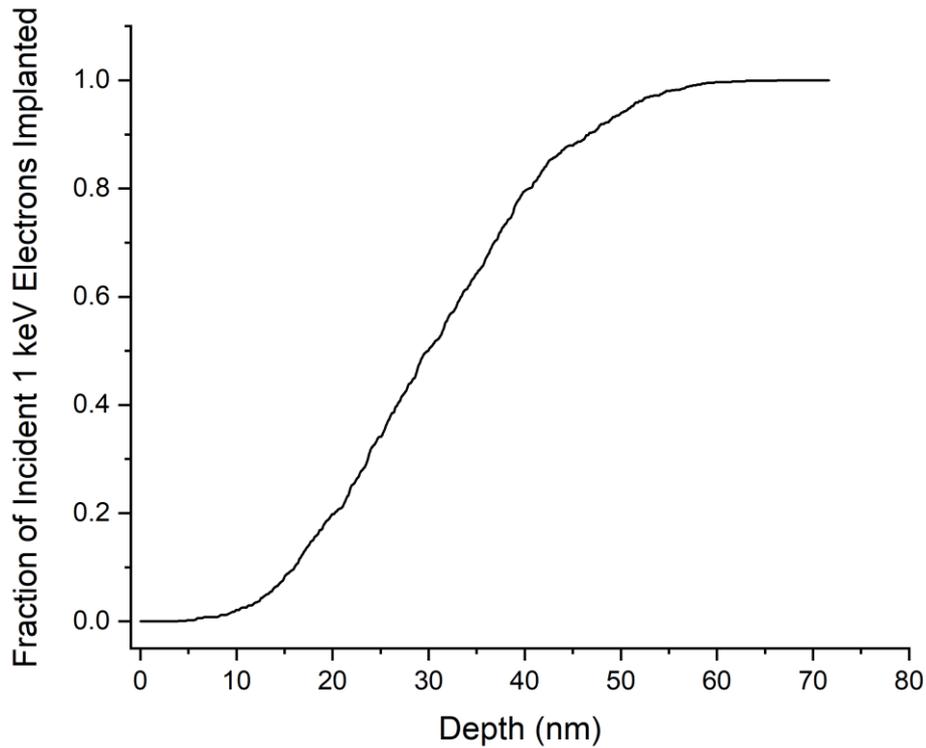

**Fig. 1:** Integrated count of 1 keV electrons implanted into the target CO$_2$ ices as a function of penetration depth as determined using the CASINO software.

## 2.2  *Experiments Performed at NSRRC*

Experimental work at NSRRC was performed using the photo-absorption and photo-luminscence end-station connected to the high-flux BL 03 beam line on the TLS: a 1.5 GeV storage ring. Detailed information on this set-up can be found in other publications (Lu et al. 2005, Kuo et al. 2007, Lu et al. 2008, Lu et al. 2013), although a brief description is provided here. VUV radiation from the high-flux beam line is dispersed using a cylindrical grating monochromator of focal length 6 m, with the radiation wavelengths having been calibrated against absorption lines for a selection of noble gases and simple molecules. This radiation is then directed towards a LiF crystal plate angled at 45° to the path of the beam line which serves as a beam splitter. The reflected portion of the radiation then passes through a LiF window before impinging onto a glass window coated with sodium salicylate. The resultant photo-luminescent signal is subsequently detected using a photo-multiplier tube set in a photon-counting mode for normalisation.

The VUV radiation transmitted through the beam splitter intersects with a cold, rotatable LiF deposition substrate onto which astrophysical ice analogues may be prepared. The radiation propagates through this substrate and impinges onto another glass window coated with sodium salicylate, from which the converted visible light is measured with a second photo-multiplier. The deposition substrate itself is connected to a cryostat able to cool the substrate to 10 K and is held at a pressure of a few $10^{-8}$ mbar. Astrophysical ice analogues are typically synthesised by first preparing gases in a pre-dosing line using standard manometric practices and subsequently directly depositing the gases onto the cold substrate at a normal angle of incidence.

The work performed at NSRRC involved the thermal annealing of pure, unirradiated $CO_2$ ices deposited at 10, 25, 30, 40, and 50 K so as to determine the effect of temperature on the acquired VUV spectra and provide complementary data to similar work performed at Atomki where mid-IR spectroscopy was utilised. To perform this work, a VUV spectrum of the bare deposition substrate was first recorded so as to be used as a reference for subsequent spectra. Deposition of pure $CO_2$ to final thicknesses of 0.015-0.022 µm was then performed at the desired temperature (i.e., 10, 25, 30, 40, or 50 K) using $CO_2$ gas (99.95% from Matheson). Once deposition was completed, another VUV spectrum was acquired to allow for absorbance values to be calculated using the Beer-Lambert Law. In all spectra, a resolution of 1 nm was used. Once deposited, ices were cooled to 10 K, re-warmed to the deposition temperature, and then gradually warmed to 70 K with spectra being acquired at 10 K intervals. For the ice deposited at 25 K, cooling to 10 K was followed by re-warming to 30 K (instead of the initial deposition temperature of 25 K), after which warming to 70 K was performed in the same manner as for the ices deposited at other temperatures.

**Table 1:** Summary of experiments performed as part of this study.

| Experiment | Deposition / Irradiation $T$ (K) | Spectroscopy Type | Laboratory |
|---|---|---|---|
| *Thermal Annealing of $CO_2$ Ices* | 25, 30, 40, 50 | Mid-IR | Atomki |
| | 10, 25, 30, 40, 50 | VUV | NSRRC |
| *Electron Irradiation of $CO_2$ Ices* | 20, 30, 40, 50 | Mid-IR | Atomki |

## 3 Results and Discussion

### 3.1 Thermal Annealing of Pure, Unirradiated $CO_2$ Ices

Infrared spectra of $CO_2$ ices deposited at 25, 30, 40, and 50 K and thermally annealed to 70 K were acquired using mid-IR absorption spectroscopy. All ices displayed a double-peaked $\nu_2$ bending mode in the infrared centred around 660 and 654 cm$^{-1}$ (15.15 and 15.29 µm) irrespective of the deposition temperature (Fig. 2). As discussed above,

Gerakines and Hudson (2015) had suggested that such a double-peaked bending mode is characteristic of $CO_2$ ices exhibiting some form of crystallinity, with amorphous structures instead exhibiting a broader single-peaked structure. However, this was disputed by Baratta and Palumbo (2017), who suggested that the appearance of a broad single-peaked bending mode was more likely to be caused by the presence of contaminant species such as $H_2O$. Given that we have not observed any traces of contamination in our spectra, nor in quadrupole mass spectrometric measurements of the gas phase of the chamber, we are inclined to state that the number of peaks present in the bending mode is not a good indication of the amorphous or crystalline nature of the ice, but rather of its purity, as suggested by previous studies (Ehrenfreund et al. 1997, van Broekhuizen et al. 2006, Baratta and Palumbo 2017).

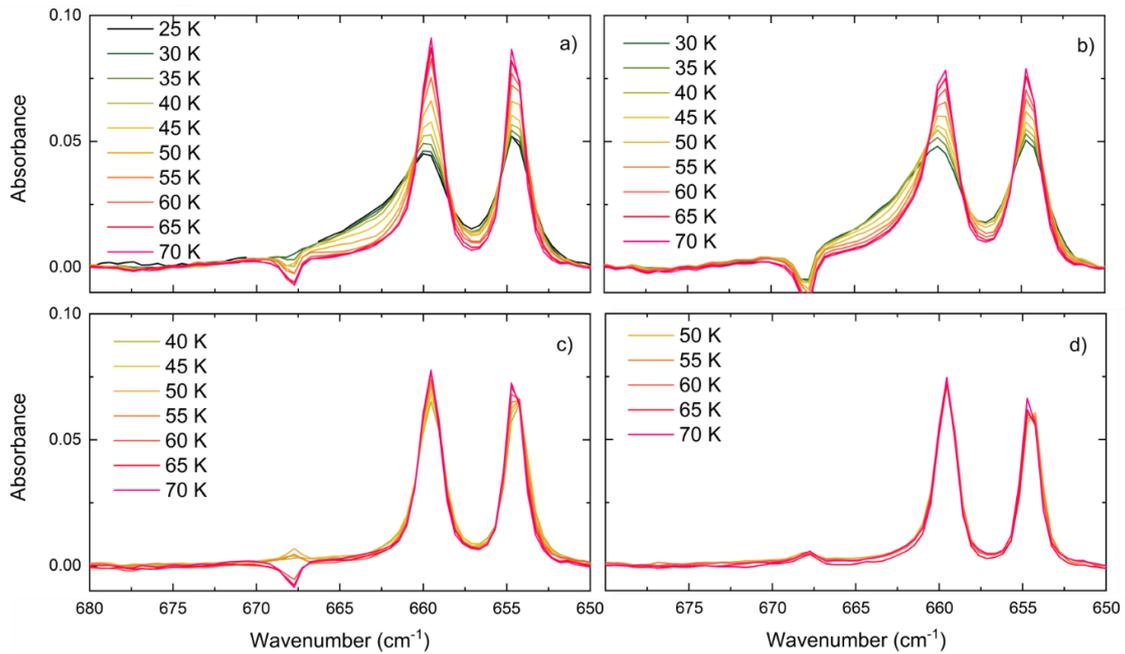

**Fig. 2:** Thermal evolution of the $CO_2$ $\nu_2$ infrared bending mode for pure, unirradiated ices deposited at (a) 25, (b) 30, (c) 40, and (d) 50 K. The high-wavenumber 'wing' structure indicative of amorphousness is clearly present in ices deposited at 25 and 30 K, but disappears during thermal annealing and is not evident at 45 K and above.

We note that our spectroscopic non-detection of contaminant species such as $H_2O$ may be explained by the high deposition pressures used that effectively excluded contaminants from the bulk ice, which is what is observed when using transmission absorption infrared spectroscopy. Additionally, the spectrometric non-detection of $H_2O$ and short experimentation times meant that the slow accretion of contaminant $H_2O$ only resulted in the deposition of, at most, a few monolayers at the surface of the $CO_2$ ice. This is not sufficient to influence our thermal annealing experiments by capping over the deposited ice, and is not expected to influence the outcome of electron irradiation, which is a bulk process.

In our work, ices deposited at 25, 30, and (to a significantly lesser extent) 40 K also displayed a broader $\nu_2$ mode accompanied by a 'wing' structure at lower temperatures (Fig. 2). Upon thermal annealing of the ices, however, this 'wing' began to disappear gradually until it could no longer be observed at 45 K and above. The disappearance of this 'wing', along with the fact that the double-peaked structure became sharper and narrower with increasing temperatures, indicated that a re-structuring process occurred within the ice; possibly a transition to a more crystalline structure characterised by a more extensive lattice (Isokoski et al. 2013). This induced re-structuring was observed to begin at a temperature above 35-40 K and was largely complete by 45 K. Indeed, the $CO_2$ ice deposited at 50 K depicted a sharp double-peaked $\nu_2$ absorption band with no broad 'wing' structure at the higher wavenumber end of the 660 cm$^{-1}$ (15.15 μm) band, suggesting a more ordered ice structure.

Additional information on the structure of the $CO_2$ ice was also gleaned from the infrared $\nu_3$ stretching mode (Fig. 3). When analysing this absorption band, three of its features must be considered: (i) the position of the band peak, (ii) the broadness of the band, and (iii) the presence or absence of a 'shoulder' band at 2328 cm$^{-1}$ (4.30 μm). Our analysis revealed that the position of the peak of the $\nu_3$ band did not shift upon warming, but rather stayed fixed at 2344 cm$^{-1}$ (4.27 μm). This lack of band peak shifting within a resolution of 1 cm$^{-1}$ was observed for $CO_2$ ices at all investigated deposition temperatures. Such a result is consistent with the work of Isokoski et al. (2013), who demonstrated using highly resolved mid-IR spectra that the maximum red-shift of the $\nu_3$ band is 0.72 cm$^{-1}$ when warming from 15 to 75 K.

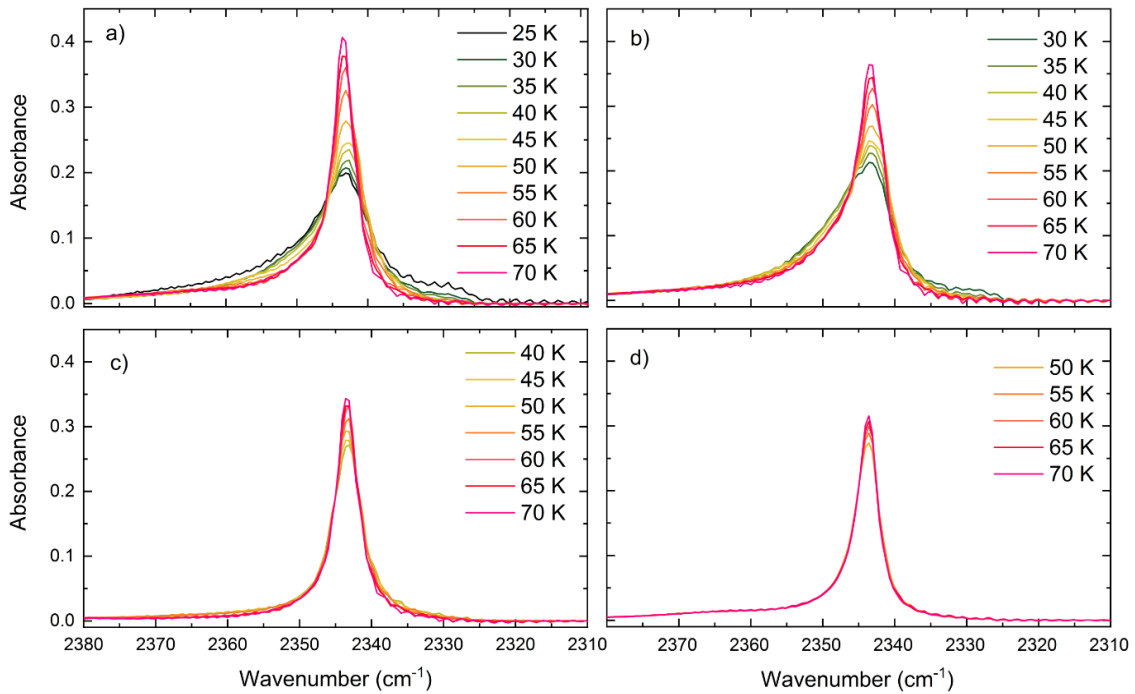

**Fig. 3:** Thermal evolution of the $CO_2$ $\nu_3$ infrared asymmetric stretching mode for pure, unirradiated ices deposited at (a) 25, (b) 30, (c) 40, and (d) 50 K. Note that no 'shoulder' band at 2328 cm$^{-1}$ (4.30 μm) was observed for any of the ices, except for a tentative detection in the ice at 25 K.

During thermal annealing, the $\nu_3$ band was also observed to become narrower indicating that a process of re-structuring to a more ordered phase was underway (Fig. 3). The classification of absorption bands as either 'broad' or 'narrow' is largely qualitative, however we have attempted to make this classification somewhat more quantitative by measuring the changes in the full-width at half-maximum (FWHM) values of the $\nu_3$ band during thermal annealing. The results, normalised to the initial FWHM upon deposition, are presented in Fig. 4. As may be seen, ices deposited at any of the investigated temperatures undergo thermally-induced re-structuring to a more ordered phase as implied by a continually decreasing FWHM value. On the basis of the data shown in Figs. 3 and 4, we suggest that the greatest changes in FWHM values have occurred by the time a temperature of 45-50 K is reached, thus further supporting our hypothesis that the transition to a more ordered structure occurs largely below this temperature.

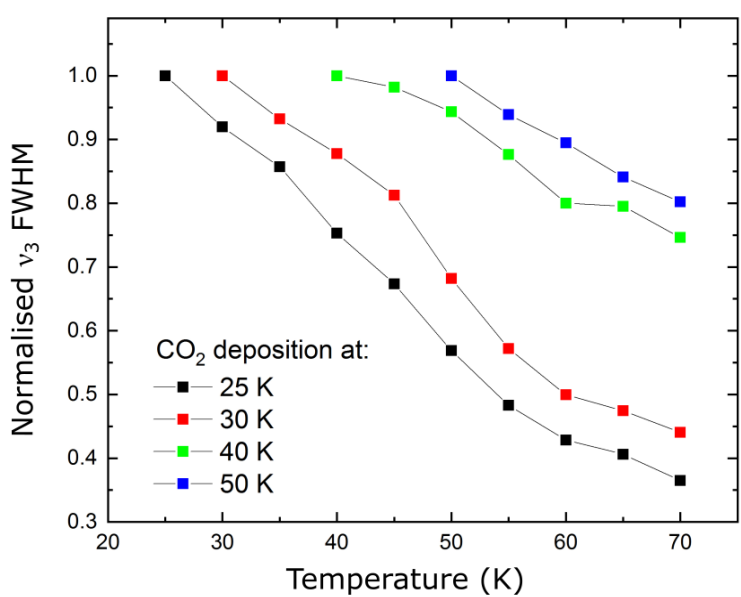

**Fig. 4:** Variations in the FWHM of the $\nu_3$ band during thermal annealing of the pure $CO_2$ ices deposited at 25, 30, 40, and 50 K. Note that, in this graph, the FWHM values have been normalised to the initial value measured upon deposition of the ice.

Perhaps one of the most interesting features of the $\nu_3$ absorption band is the presence or absence of a 'shoulder' band at 2328 cm$^{-1}$ (4.30 μm) characteristic of pure amorphous $CO_2$ as described by Escribano et al. (2013). In our experiments, we have only made a very tentative detection of this feature in the ice at 25 K prior to any thermal annealing. This 'shoulder' band could not be observed upon warming this ice to higher temperatures nor in any of the ices deposited at higher temperatures. Intuitively, this would suggest that, other than the ice at 25 K, our ices were not amorphous. However, we note that Escribano et al. (2013) reported that this 'shoulder' band is more apparent in pure amorphous ices which are only a few nanometres thick and that their spectra show a sharp decline in the peak height of this band after a few hundred monolayers of the ice are deposited. Given that our ices are much thicker than those considered by Escribano et al. (2013), the presence or absence of this 'shoulder' band is likely not particularly useful when assessing the extent of structural order. Other mid-IR spectral features, such as the broadness of the $\nu_2$ and $\nu_3$ bands and the presence of a 'wing' structure on the former may be more appropriate. Terahertz spectroscopy may also be more categorical in determining whether a $CO_2$ ice is amorphous or crystalline (Allodi et al. 2014, McGuire et al. 2016, Mifsud et al. 2021b).

Electronic spectra of thermally annealed pure $CO_2$ astrophysical ice analogues were also acquired in the VUV range (Fig. 5). Ices deposited at 10, 25, 30, 40, and 50 K were first cooled to 10 K and subsequently warmed to 70 K as described in the methodology section. Our analysis focused on the absorption bands centred at 125 nm (9.9 eV) and 141 nm (8.8 eV), corresponding to the $^1\Pi_g \leftarrow {}^1\Sigma_g^+$ and the $^1\Delta_u \leftarrow {}^1\Sigma_g^+$ transitions, respectively. The shape and profile of these bands visibly change under the influence of changing temperature, allowing for the physico-chemical effects of thermal annealing to be discerned.

For ices deposited at 10 and 25 K, the higher wavelength band at 143 nm presents as a broad but distinct peak. However, upon thermal annealing of these ices to temperatures of 30 K and above, this peak appears to be suppressed to lower absorbance values (Fig. 5). This suppression was also observed for the ices deposited at 30, 40 and 50 K upon both initial deposition as well as thermal annealing to higher temperatures. The effect is most clearly observed in the case of the $CO_2$ ice deposited at 10 K, where the $^1\Delta_u \leftarrow {}^1\Sigma_g^+$ band is initially visible as a broad peak whose absorbance actually exceeds that of the $^1\Pi_g \leftarrow {}^1\Sigma_g^+$ absorption band (Fig. 5). Upon annealing, however, the absorbance of the former band begins to decrease and, from 40 K onwards, appears as a signal with a flatter profile. Interestingly, when ices which had been deposited at 30, 40, and 50 K were initially cooled to 10 K prior to thermal annealing, a slight increase in the absorbance of the $^1\Delta_u \leftarrow {}^1\Sigma_g^+$ band was recorded, although not to the same extent as in the ice initially deposited at 10 K (Fig. 5).

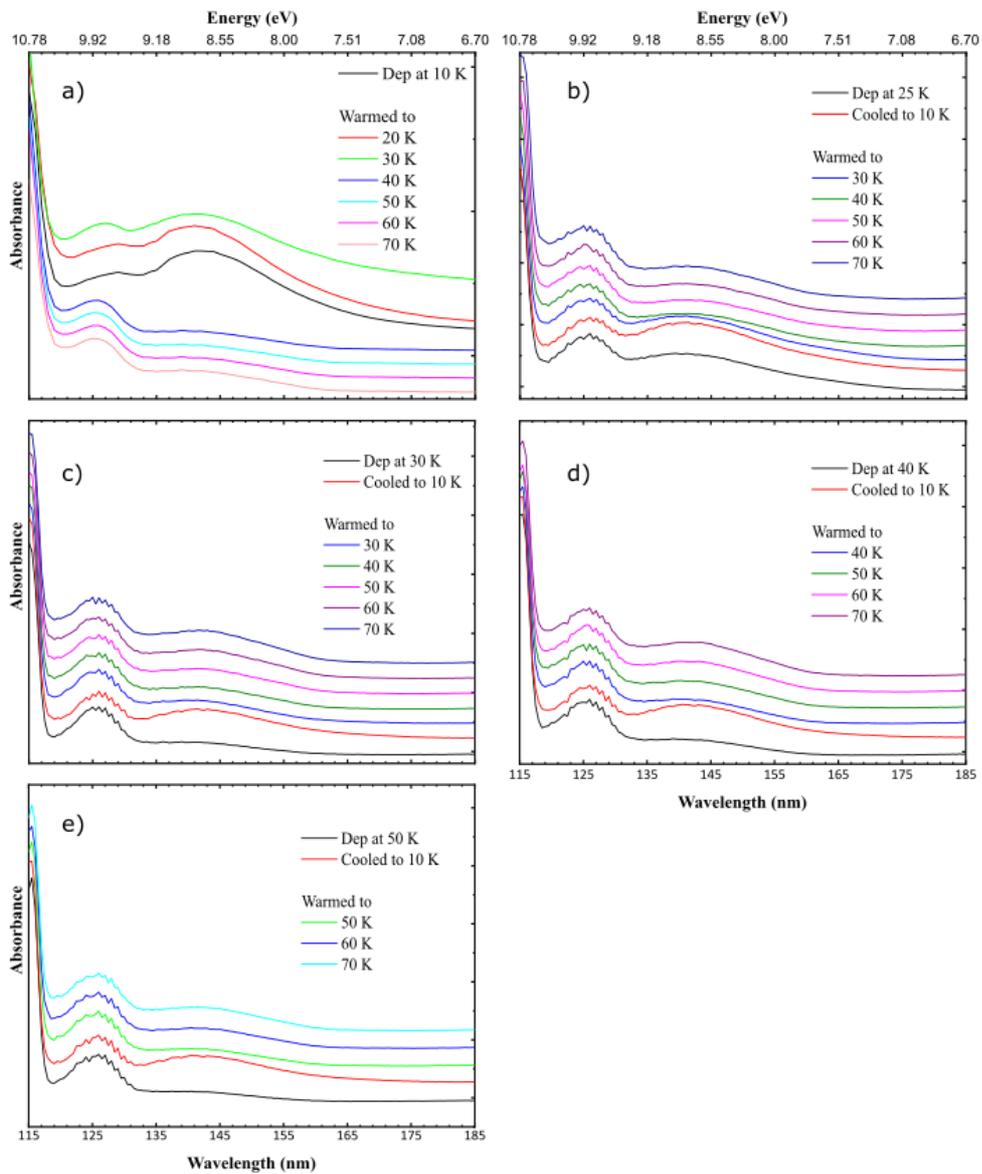

**Fig. 5:** Thermal evolution of the VUV spectrum of pure $CO_2$ in the region 115-185 nm (10.8-6.7 eV) for ices deposited at (a) 10, (b) 25, (c) 30, (d) 40, and (e) 50 K on annealing to 70 K.

Such results are suggestive of a re-structuring process occurring within the ice. These findings in the VUV complement well our mid-IR data which suggest a thermally-induced transition towards a more ordered structure with a more extensive lattice occurring in the temperature range 35-45 K, as evidenced by the $\nu_2$ and $\nu_3$ bands becoming sharper as well as the gradual loss of the 'wing' structure on the high-wavenumber end of the former band.

The findings of this spectral study are directly relevant to astrophysics, both in the context of interstellar chemistry as well as that of planetary and lunar science. Previous studies have already established that $CO_2$ ices adsorbed to interstellar dust grains are ordered structures, having adopted some degree of crystallinity as a result of segregation from polar ice layers or the distillation of volatile apolar ice mixture constituents such as CO or $O_2$ (Escribano et al. 2013, He et al. 2017, Tsuge et al. 2020, Kouchi et al. 2021). Our results indicate that mid-IR and VUV spectra are indeed sensitive to structural changes induced by thermal annealing, but may not be as categorical in defining a $CO_2$ ice as amorphous or crystalline as other techniques, such as terahertz spectroscopy (Allodi et al. 2014, McGuire et al. 2016, Mifsud et al. 2021b). Nevertheless, key spectral features, such as the absence of the 'wing' structure in the mid-IR $\nu_2$ bending mode and the suppression of the $^1\Delta_u \leftarrow {}^1\Sigma_g^+$ band in the VUV, provide good indications of the extent of ice re-structuring.

In the context of planetary science, our results suggest that $CO_2$ ice at the surfaces of icy outer Solar System moons (which are characterised by temperatures greater than 50 K) should exist as an ordered phase; possibly as crystalline or polycrystalline ices. Observations made using the Near-Infrared Mapping Spectrometer (NIMS) instrument aboard the *Galileo* spacecraft have definitively confirmed the presence of $CO_2$ as a minor surface constituent of Europa, Ganymede, and Callisto: the icy Galilean moons of Jupiter

(Hibbitts et al. 2000, Hibbitts et al. 2003, Hansen and McCord 2008). Although there is, to the best of the authors' knowledge, no study specifically looking into the phase of this surface $CO_2$ ice, it is indeed thought that the $CO_2$ on these moons is largely crystalline (Delitsky and Lane 1998), although some $CO_2$ is also believed to be present as inclusions within clathrate hydrates (Prieto-Ballesteros et al. 2005, Oancea et al. 2012).

### 3.2    1 keV Electron Irradiation of Pure $CO_2$ Ices

Pure $CO_2$ astrophysical ice analogues were irradiated using 1 keV electrons at four different temperatures: 20, 30, 40, and 50 K. Mid-IR absorption spectra collected at various times during irradiation allowed for electron-induced physico-chemical effects to be discerned. Aside from the decay of the $CO_2$ absorption bands, the appearance of bands attributable to other molecules synthesised as a result of irradiation were observed (Fig. 6, Table 2). The focuses of our study were the radiolytic product molecules CO, $CO_3$, and $O_3$.

The formation of these product molecules from $CO_2$ ices as a result of radiolysis or photolysis has been well documented in the literature (e.g., Hudson and Moore 2001, Bennett et al. 2010, Sivaraman et al. 2013, Fillion et al. 2014, Jones et al. 2014a, Jones et al. 2014b, Mejía et al. 2015, Martín-Doménech et al. 2015, Sie et al. 2019). Mechanistic studies using isotopologues have shown that, when $CO_2$ is irradiated using high-energy electrons, as was the case in our study, the first reaction is molecular dissociation resulting in the formation of CO and an oxygen atom (Anbar and Perlstein 1966, Bennett et al. 2010), as shown in Eq. 4. These oxygen atoms are typically produced with kinetic energies of a few eV (1 eV ≈ 11,605 K) which classifies them as supra-thermal. As such, this excess energy may be used to overcome reaction activation energy barriers or access otherwise unavailable excited states (Armentrout 1991, Cosby 1993a, Cosby 1993b, Bennett et al. 2010).

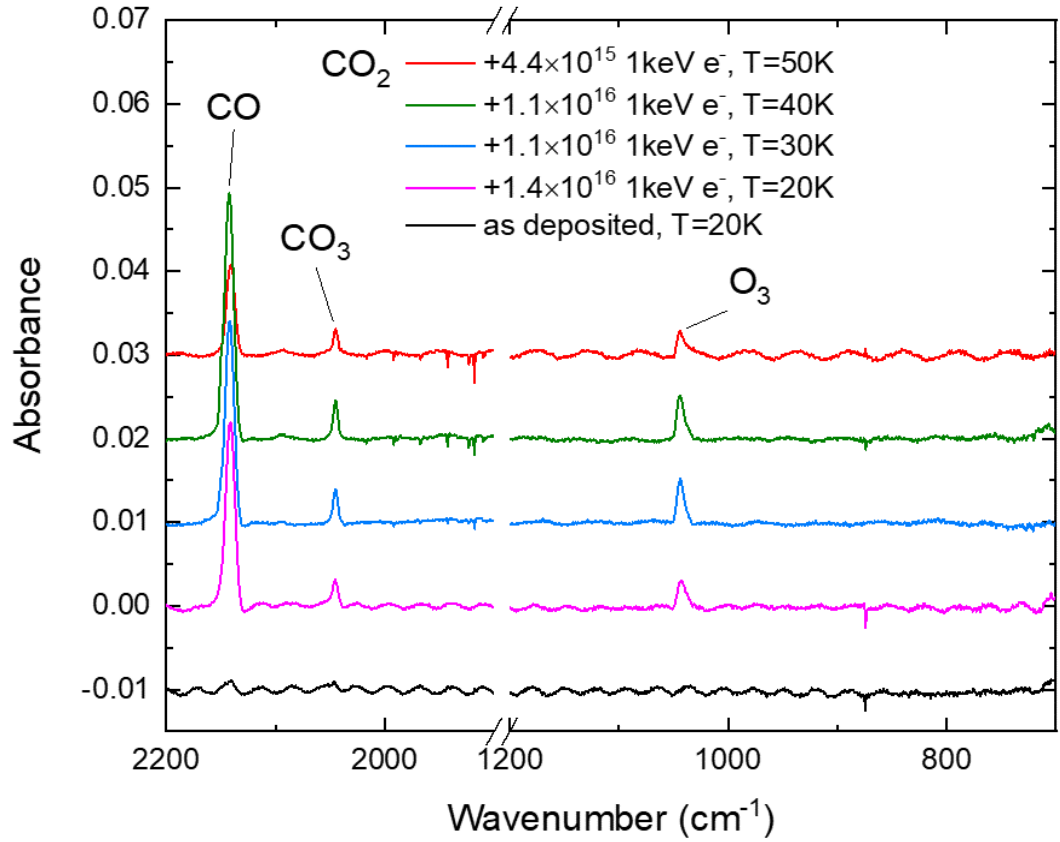

**Fig. 6:** Mid-IR absorption spectra of 1 keV electron irradiated $CO_2$ ices at 20, 30, 40, and 50 K. The irradiation resulted in the formation of CO, $CO_3$, and $O_3$, as indicated. Note that spectra are vertically offset for clarity.

**Table 2:** Mid-IR spectroscopic reference data used to identify and quantify products of $CO_2$ ice irradiation.

| Molecule | Absorption Band | $A_v$ ($10^{-17}$ cm molecule$^{-1}$) | Reference |
|---|---|---|---|
| $CO_2$ | $v_3$ (2342 cm$^{-1}$; 4.27 μm) | 7.6 | Gerakines et al. (1995) |
| CO | $v_3$ (2141 cm$^{-1}$; 4.67 μm) | 1.1 | Gerakines et al. (1995) |
| $CO_3$ | $v_1$ (2044 cm$^{-1}$; 4.89 μm) | 1.5 | Martín-Doménech et al. (2015) |
| $O_3$ | $v_3$ (1044 cm$^{-1}$; 9.58 μm) | 1.4 | Smith et al. (1985) |

The addition of such atoms to $CO_2$ results in the formation of $CO_3$ (Eq. 5). Previous studies have demonstrated that both the cyclic ($C_{2v}$) and acyclic ($D_{3h}$) structural isomers of $CO_3$ may be formed via the irradiation of $CO_2$ ice (Jamieson et al. 2006, Bennett et al. 2010, Sivaraman et al. 2013). In our study, however, we have only been able to identify the $C_{2v}$ isomer through its $\nu_1$ stretching mode at 2044 cm$^{-1}$ (4.89 μm). This may be due to the fact that this isomer is in fact the more stable one (Jamieson et al. 2006) and, given that $CO_3$ is the least abundant molecular product within the ice matrix, it is possible that any $D_{3h}$ isomer produced was not formed in a sufficient quantity as to be detected by our spectroscopic instruments.

The final product under consideration, $O_3$, is formed via a two-step reaction mechanism which relies first on the accumulation of sufficient $O_2$ as a result of the combination of supra-thermal oxygen atoms (Eq. 6). A subsequent addition of atomic oxygen then furnishes $O_3$ (Eq. 7). Previous studies have considered this latter reaction to be either energetically barrierless (Bennett and Kaiser 2005, Sivaraman et al. 2007), or else as having a small activation energy barrier of <0.05 eV (Ioppolo et al. 2008).

$$CO_2 \rightarrow CO + O \tag{Eq. 4}$$

$$CO_2 + O \rightarrow CO_3 \tag{Eq. 5}$$

$$2\,O \rightarrow O_2 \tag{Eq. 6}$$

$$O + O_2 \rightarrow O_3 \tag{Eq. 7}$$

Although such radiation chemistry is well documented in the literature, considerably fewer studies have looked into the influence of temperature on the formation of these products. One notable study was that of Sivaraman et al. (2007), which looked into the formation of $O_3$ as a result of the electron irradiation of $O_2$ ices at various temperatures. That study found that, although the addition of a supra-thermal oxygen atom to $O_2$ is a temperature-independent reaction, the loss of available atoms to do so due to their recombination to reform $O_2$ is greater at higher temperatures, thus making $O_3$ formation less efficient. Upon post-irradiative thermal annealing of the ices, however, $O_3$ formation increased due to the reaction of oxygen atoms trapped within the ice matrix with $O_2$ molecules, or due to the reaction of $O_2$ molecules with $[O_3…O]$ complexes thus synthesising $[O_3…O_3]$ complexes (Sivaraman et al. 2007).

Fig. 7 exhibits the results of our systematic electron irradiations at 20, 30, 40, and 50 K by depicting the variation in the molecular column density of $CO_2$, CO, $CO_3$, and $O_3$ with increasing projectile electron fluence. As can be seen, the decay of $CO_2$ in the ice is slowest at 20 K and becomes more rapid with increasing temperature. Such a result is in line with other studies concerning the irradiation of pure ices (e.g., Orlando and Sieger 2003, Teolis et al. 2009). At higher temperatures, $CO_2$ molecules, possibly arranged as a dimeric structure (Sivaraman et al. 2013), are less bound to their neighbours and therefore find it more difficult to disperse any excess energy which increases their propensity to undergo dissociation.

The net result of such an increased rate of molecular dissociation is, naturally, an increase in the formation of CO and supra-thermal oxygen atoms. Indeed, the accumulation of CO was noted to be more rapid at 40 K than at 30 K which, in turn, was more rapid than that at 20 K (Fig. 7). However, under extended irradiation, the molecular column density for the CO produced at 30 and 40 K began to decline, as opposed to that at 20 K which still increased further. This may have been due to two processes. Firstly, the greater abundance of supra-thermal oxygen atoms produced as a result of $CO_2$ dissociation at these higher temperatures could have promoted the recombination reaction leading back to $CO_2$ (Eq. 8), as the number of successful collisions would increase due to the greater abundance of CO within the ice matrix.

$$CO + O \rightarrow CO_2$$

(Eq. 8)

This, however, is unlikely, as previous studies have demonstrated that this reaction is not a particularly favourable one (Roser et al. 2001, Madzunkov et al. 2006, Raut and Baragiola 2011, Ioppolo et al. 2013). Alternatively, it is possible that the physico-chemical effects induced by such prolonged irradiation may have created defects within the ice such as cracks or pores, through which CO may have sublimated into the gas phase. Sublimation of CO is, in fact expected at temperatures above ~30 K, but may be shifted to higher temperatures if the CO is trapped within the ice matrix (Hudson and Donn 1991, Fray and Schmitt 2009). As such, the comparatively low abundance of CO observed in the ice via mid-IR spectroscopy during the irradiation at 50 K (which peaks at about two-and-a-half times less than that at 40 K) is attributed to a larger scale sublimation as CO escapes through radiation-induced ice defects.

While CO is the most abundant of the radiolytic product molecules, $CO_3$ is the least abundant. This implies that the addition of a supra-thermal oxygen atom to $CO_2$ is a less efficient process than the two-step production mechanism of $O_3$, with typical $O_3$ column densities being twice those for $CO_3$ for a given temperature and electron fluence. Overall, however, the temperature dependence of $CO_3$ synthesis is fairly similar to that for CO, with production peaking at a higher abundance at 40 K than at 30 K, which in turn is higher than that at 20 K (Fig. 7). This, once again, may be due to an increase in the production of supra-thermal oxygen atoms released by $CO_2$ dissociation, and such a result is also in line with that of Sivaraman et al. (2013) who also proposed a second formation mechanism for $CO_3$ based on a dimeric $CO_2$ complex:

$$[CO_2…CO_2] \rightarrow [CO…O…CO_2] \rightarrow [CO…CO_3] \rightarrow CO + CO_3$$

(Eq. 9)

An interesting consideration is also the formation of $CO_3$ as a result of the direct reaction between CO and $O_2$ (Eq. 10); the latter being formed via the combination of two oxygen atoms (Eq. 6). Previous work by Jamieson et al. (2006) has highlighted that, although this reaction is possible and should be invoked to explain observed molecular column densities (particularly in the case of the $C_{2v}$ isomer), it is not efficient and the role of $O_2$ in the formation of $CO_3$ is likely to be minimal. The production of $CO_3$ as a result of $CO_2$ irradiation at 50 K was observed to be significantly lower than at other, lower temperatures; as was the case for CO. However, in this case, we consider the desorption or sublimation of this molecule from the bulk ice to be unlikely, as studies investigating the photon irradiation of $CO_2$ ices have not been successful in detecting gas-phase $CO_3$, despite detecting other photochemical products (Bahr and Baragiola 2012, Martín-Doménech et al. 2015). Instead, we suggest that this depletion may be due to the increased rate constants for $CO_3$ dissociation at higher temperatures (Eq. 11), and indeed $CO_3$ molecular column densities are observed to decrease at earlier electron fluences and to lower values as the irradiation temperature is raised (Fig. 7).

$$CO + O_2 \rightarrow CO_3$$

(Eq. 10)

$$CO_3 \rightarrow CO_2 + O$$

(Eq. 11)

Finally, we consider the influence of temperature on the evolution of $O_3$. Although other studies have considered the formation of $O_3$ as a result of photon or charged particle processing of $CO_2$ ices (Ennis et al. 2011, Sivaraman et al. 2013, Mejía et al. 2015, Martín-Doménech et al. 2015), to the best of our knowledge only that of Sivaraman et al.

(2013) has considered more than one temperature. That study found that the abundance of $O_3$ produced as a result of $H^+$, $D^+$, or $He^+$ irradiation at 30 K was three to five times higher than that produced at 80 K, depending on the projectile used.

Upon first inspection, it is perhaps intuitive to assume that this result contrasts or contradicts the data presented in Fig. 7, where $O_3$ production is seen to increase significantly on raising the irradiation temperature from 20 to 30 K, and then increase much more modestly on going from 30 to 40 K. This therefore implies a positive correlation between temperature and $O_3$ radiolytic synthesis. However, our results also show that $O_3$ production at 50 K results in a much lower abundance of the molecule in the ice matrix compared to that at lower temperatures. This is expected, as previous work has shown that the precursor molecules to $O_3$, such as $O_2$, desorb efficiently in the temperature range 30-40 K (Collings et al. 2004, Fray and Schmitt 2009). In an ice matrix of $CO_2$, this desorption temperature is increased and our data suggests that sublimation seems to occur between 40-50 K, thus explaining the lower abundance of $O_3$ observed at 50 K. Additionally, $O_3$ itself sublimates in the 61-68 K temperature range (Chaabouni et al. 2000), meaning that some desorption of the product molecule at 50 K through radiation-induced ice defects is also possible, although this is likely not as much of a contributing factor to the low $O_3$ abundance at 50 K as the sublimation of the precursor molecules. Similar arguments were invoked to explain the 80 K results of Sivaraman et al. (2013), thus reconciling our results with that study.

As mentioned earlier, previous work on the irradiation of pure $O_2$ ice highlighted the fact that higher temperatures were actually associated with lower $O_3$ production rates due to more efficient recombination of supra-thermal oxygen atoms to recycle $O_2$ (Sivaraman et al. 2007). Such a result, however, is not observed on changing the irradiated ice target to $CO_2$, as is evident from the results of this present study. This is to be expected, as the number of reaction pathways available to these atoms is greater when they are yielded from $CO_2$ (Eqs. 5-9) compared to when they are sourced from $O_2$ (Eqs. 6 and 7), leading to more productive overall chemistry. Additionally, the recombination reaction between a supra-thermal oxygen atom and CO leading back to $CO_2$ (Eq. 8) has been shown to be a comparatively inefficient process (Roser et al. 2001, Madzunkov et al. 2006, Raut and Baragiola 2011, Ioppolo et al. 2013), and so is not as influential as the combination of two oxygen atoms to yield $O_2$ (Eq. 6).

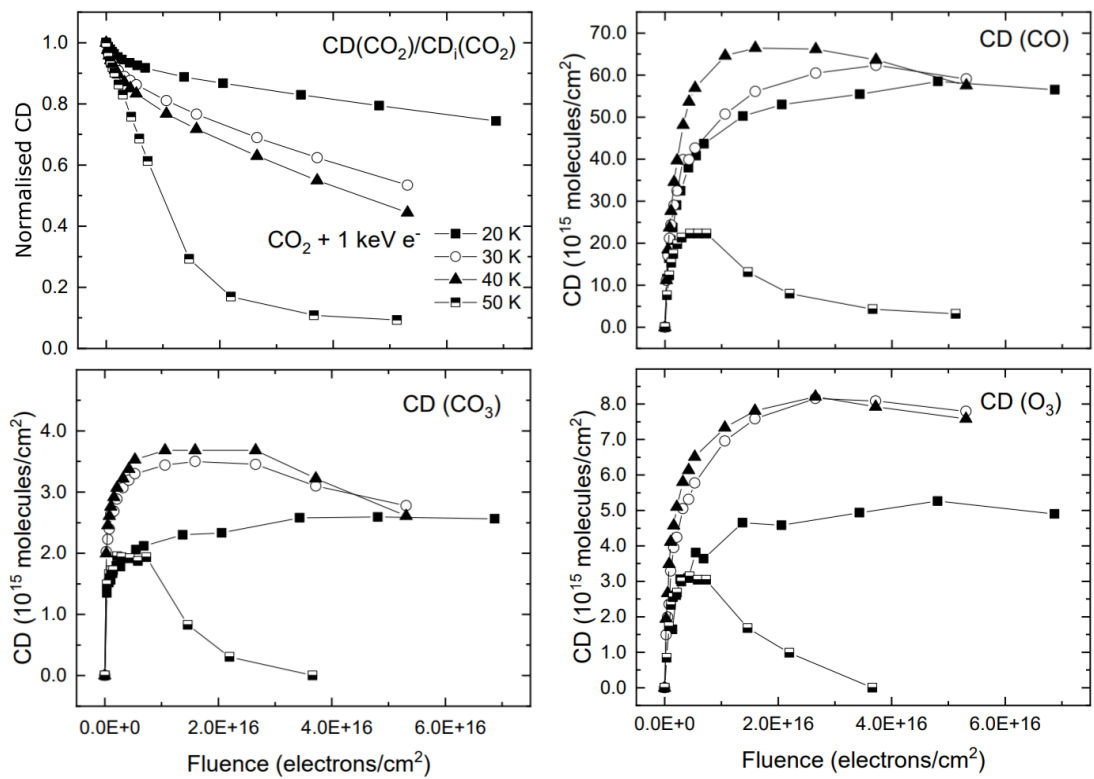

**Fig. 7:** Plots of molecular column density for $CO_2$, CO, $CO_3$, and $O_3$ against 1 keV electron fluence. Note that, for $CO_2$ the molecular column density has been normalised to the initial column density measured upon deposition of the ice. Furthermore, in this figure, the notation 'CD' refers to column density, while '$CD_i$' refers to the column density of the ice which was initially deposited.

## 4      Summary and Concluding Remarks

We have carried out a systematic analysis of the influence of temperature on two aspects of $CO_2$ astrochemistry: (i) its influence on the shape and appearance of the mid-IR and

VUV absorption spectra of pure, unirradiated $CO_2$ ices, and (ii) its influence on the production of daughter molecules (particularly CO, $CO_3$, and $O_3$) as a result of irradiation by 1 keV electrons. Our results indicate that the mid-IR and VUV absorption spectra of $CO_2$ deposited at 20, 30, 40, and 50 K and subsequently warmed to 70 K are sensitive to changes in temperature, and we have been able to quantify changes in major absorption band peaks as a result of thermal annealing (Fig. 4). Acquired spectra indicate that our $CO_2$ ices underwent progressive structural ordering as a result of thermal annealing. This is in agreement with the results of prior studies which have not only observed crystalline $CO_2$ within interstellar icy grain mantles, but have also demonstrated its resistance to amorphization under the influence of radiation (Hudson and Moore 1995, Escribano et al. 2013, He et al. 2017, Tsuge et al. 2020).

Our 1 keV electron irradiations of pure $CO_2$ ices at 20, 30, 40, and 50 K resulted in the formation of several radiolytic product molecules, including CO, $CO_3$, and $O_3$ in agreement with previous research (Ennis et al. 2011, Sivaraman et al. 2013, Mejía et al. 2015, Martín-Doménech et al. 2015). Overall, our results indicate that molecular product formation increases noticeably on raising the irradiation temperature from 20 to 30 K as a result of increased atom, radical, and molecule mobility within the ice matrix, and then increases more modestly on further raising the temperature to 40 K (Fig. 7). At 50 K, the overall abundances of these molecules within the ice matrix were noted to be less than what had been observed at 20 K. In the cases of CO and $O_3$ we have attributed this to an increased rate of sublimation of the molecules and their precursors, while in the case of $CO_3$ we have suggested that this is the result of increased destruction by incident ionising radiation. These results constitute, to the best of our knowledge, the first systematic analysis of the temperature-dependence of $CO_2$ radiolytic chemistry.

Both the thermal annealing and radiolytic chemistry work presented in this paper are directly relevant to the astrochemistry of interstellar and Solar System ices, of which $CO_2$ is a ubiquitous component. Such ices are routinely processed by various energetic sources, including galactic cosmic rays, the solar wind, or planetary magnetospheric plasmas. As such, our spectral characterisations and identifications may prove useful by acting as reference data for forthcoming missions aimed at defining the composition and chemistry of space more accurately. The ESA *Jupiter Icy Moons Explorer* (JUICE) and the NASA *Europa Clipper* missions, in particular, will aim to further characterise the Jovian Galilean moon system due to the status of these satellites as potential abodes of extra-terrestrial life (Grasset et al. 2013, Phillips and Pappalardo 2014). Given the presence of $CO_2$ at the surfaces of these icy moons, as well as the temperature fluctuations experienced at the surface as their host planet orbits the sun, our analysis of the temperature-induced variations in spectroscopy and radiation chemistry are appropriate.

The work presented in this paper will also, hopefully, serve as a starting point for other studies seeking to further elucidate $CO_2$ ice astrochemistry. As the scientific discipline of astrochemistry expands and develops, there is a growing need for laboratory experiments to adopt a more systematic approach in which the influence of certain variables on the overall chemistry being observed is better constrained, so as to be able to replicate and understand more fully the exact mechanisms occurring in space. We intend to build on the results presented in this study by performing similar systematic studies on more complex ice mixtures containing $CO_2$ and other simple molecules of astrophysical interest in the near future.

**Acknowledgements**

The authors all acknowledge funding from the Europlanet 2024 RI which has received funding from the European Union Horizon 2020 Research Innovation Programme under grant agreement No. 871149. The main components of the ICA set-up were purchased with funding from the Royal Society through grants UF130409, RGF/EA/180306, and URF/R/191018. Support has also been received from the Hungarian Scientific Research Council Fund through grant No. K128621.

D.V. Mifsud is the grateful recipient of a University of Kent Vice-Chancellor's Research Scholarship. The research of Z. Kaňuchová is supported by VEGA – the Slovak Grant Agency for Science (grant No. 2/0023/18) and the Slovak Research and Development Agency (contract No. APVV-19-0072). S. Ioppolo gratefully acknowledges the Royal Society for financial support. A. Traspas Muiña thanks Queen Mary University of London for doctoral funding. The research of B. Paripás is supported by the European Union and the State of Hungary and co-financed by the European Regional Development Fund (grant GINOP-2.3.4-15-2016-00004).

S. Pavithraa, K.K. Rahul, A. Das, B.N. Raja Sekhar, and B. Sivaraman thank the Sir John and Lady Mason Academic Trust for support to perform experiments at NSRRC. Furthermore, A. Bhardwaj and B. Sivaraman would like to thank the Physical Research Laboratory (Department of Space, Government of India) for the support which enabled this work to be carried out. B. Sivaraman also acknowledges support from INSPIRE (grant IFA-11 CH-11) which enabled much of the VUV spectroscopy described in this study to be performed. A. Das acknowledges support from the Grant-In-Aid from the Higher Education Department of the Government of West Bengal. The research of B.-M. Cheng is supported by the Ministry of Science and Technology of Taiwan (grant No. 108-2113-M-213-003). We also wish to express our thanks to NSRRC for providing the facility where our VUV absorption measurements were conducted.

**Graphical Abstract**

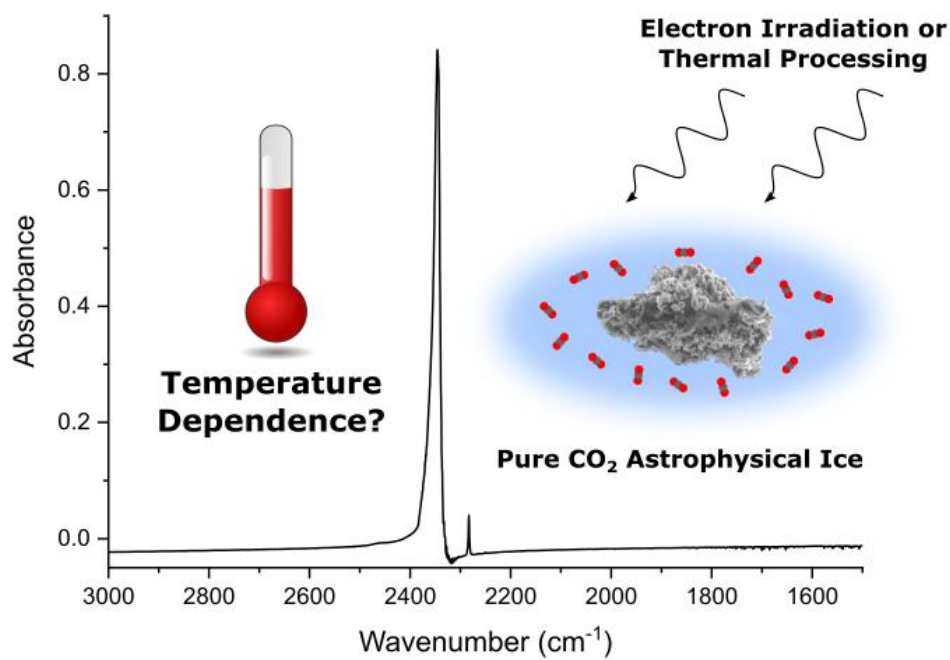

**Conflict of Interests Statement**

The authors hereby state that they have no known conflicts of interest which may have biased the outcome or interpretation of the results presented in this study.